*Research Article*

# A Slice-Based Change Impact Analysis for Regression Test Case Prioritization of Object-Oriented Programs

**S. Panda, D. Munjal, and D. P. Mohapatra**

*Department of Computer Science and Engineering, National Institute of Technology, Rourkela, Sundergarh, Odisha 769008, India*

Correspondence should be addressed to S. Panda; subhrakanta11@gmail.com

Test case prioritization focuses on finding a suitable order of execution of the test cases in a test suite to meet some performance goals like detecting faults early. It is likely that some test cases execute the program parts that are more prone to errors and will detect more errors if executed early during the testing process. Finding an optimal order of execution for the selected regression test cases saves time and cost of retesting. This paper presents a static approach to prioritizing the test cases by computing the affected component coupling (ACC) of the affected parts of object-oriented programs. We construct a graph named affected slice graph (ASG) to represent these affected program parts. We determine the fault-proneness of the nodes of ASG by computing their respective ACC values. We assign higher priority to those test cases that cover the nodes with higher ACC values. Our analysis with mutation faults shows that the test cases executing the fault-prone program parts have a higher chance to reveal faults earlier than other test cases in the test suite. The result obtained from seven case studies justifies that our approach is feasible and gives acceptable performance in comparison to some existing techniques.

## 1. Introduction

In the software development life cycle, regression testing is considered an important part. This is because it is essential to validate the modification and to ensure that no other parts of the program have been affected by the change [1]. Regression testing [2–6] is defined as the selective retesting of a system or component to verify that modifications have not caused unintended effects and that the system or component still complies with its specified requirements [2]. Therefore, this paper follows a selective approach [3–5] to identify and retest only those parts of the program that are affected by the change. Thus, it is even more important to improve the effectiveness of regression testing and reduce the cost of test case execution. Therefore, in this paper, we focus on test case prioritization (TCP) of a given test suite $T$. Test case prioritization problem is best described using the example in Table 1. If the test cases are executed in the order $\{T1, T2, T3, T4, T5, T6\}$, then we achieve 100% coverage of faults only after the sixth test case is executed, whereas if the ordering of the test case execution is changed to $\{T6, T5, T4, T1, T3, T2\}$, then we achieve 100% coverage after the execution of the fourth test case. Therefore, finding the order of test case execution is essential to detect the faults [7–9] early during regression testing. All the existing approaches target finding an optimal ordering of the test cases based on the rate of fault detection or rate of satisfiability of coverage criterion under consideration. However, these existing techniques [3, 6, 10–13] were primarily developed to target procedural programs. Very few existing works [14–18] focus on the test case prioritization for object-oriented programs. This paper presents a static approach of prioritizing the test cases of object-oriented programs. It is reported that a module having high coupling value is more erroneous than other modules [19, 20]. As a result, a test case that executes a module with high coupling value will reveal more faults than other test cases. Many techniques and metrics [21] exist to measure the coupling value of the program segments [19] and establish these values as an indicator of fault-proneness [20]. None of the prioritization techniques available in the literature have reported the use of coupling measure to prioritize the test cases. Thus, this paper uses the coupling value of the affected program parts for prioritizing the selected test cases for regression testing.

Table 1: A sample test case distribution and the faults detected by them.

| Test cases/faults | $T1$ | $T2$ | $T3$ | $T4$ | $T5$ | $T6$ |
|---|---|---|---|---|---|---|
| $f1$ | X | | | | X | X |
| $f2$ | | | X | X | | X |
| $f3$ | X | | X | | X | |
| $f4$ | | | X | | | X |
| $f5$ | | X | | X | | |
| $f6$ | | | | | X | X |
| $f7$ | | | | | | X |
| $f8$ | X | | | | | |
| Number of faults | 3 | 1 | 3 | 2 | 3 | 5 |
| % of faults detected by two sample test case orderings | | | | | | |
| $T1, T2, T3, T4, T5, T6$ | 37.5 | 50 | 75.0 | 75.0 | 87.5 | 100 |
| $T6, T5, T4, T1, T3, T2$ | 62.5 | 75 | 87.5 | 100 | 100 | 100 |

Based on the above motivations, we propose an approach to prioritize a selected test suite of an object-oriented program using the coupling value of the affected program parts covered by the test cases. For experimentation, we have taken a sample Java program shown in Algorithm 1. A total of twenty test cases ($T1$–$T20$) were taken along with their node coverage information. All those test cases that covered the affected nodes (with respect to a modification point) are selected hierarchically. Finally, five test cases ($T6$–$T10$) are selected for prioritization. For hierarchical regression test case selection details interested readers are requested to refer to [19]. In this approach, we propose a technique to prioritize the selected test cases ($T6$–$T10$). Thus, we fix our research objectives as follows:

(i) To identify and represent the affected program parts and compute the coupling value of these affected program parts.

(ii) To cluster the coupling values [22] into groups and assign a weight to each group based on their criticality.

(iii) To prioritize the test cases by sorting them in the decreasing order of their computed weights.

So, the contributions of this paper lies in the following:

(i) Proposing an algorithm for prioritizing the selected test cases.

(ii) Implementing the proposed algorithm for the fifteen experimental programs.

(iii) Carrying out mutation analysis.

(iv) Comparing the performance of our approach with an existing work.

The rest of the paper is organized as follows: Section 2 introduces the technique used in this paper for prioritizing the test cases. We describe our proposed process of prioritization in Section 3. We also discuss the working and complexity analysis of our algorithm in this section. The details of our implementation and experimental studies are given in Section 4. Here, we present the experimental study settings, describe the characteristics of the program samples taken for our experimentation and mutation analysis, and analyze the results. In Section 5, we discuss and compare our work with some related work. We also highlight some of the limitations of our approach in this section. We conclude the paper in Section 6 with some insights into probable extensions to our work.

## 2. Preliminary Study

In this section, we discuss the techniques that are used in this work to accomplish our research objectives.

*2.1. Program Slicing.* This paper uses program slicing to identify the affected program parts for change impact analysis. Program slicing was originally introduced by Weiser [23] as a method for automatically decomposing programs by analyzing their data flow and control flow starting from a subset of a program's behavior. Slicing reduces the program to a minimal form that still produces the same behavior as the original program. Program slicing is a method of separating out the relevant parts of a program with respect to a particular computation [24–27]. The input that the slicing algorithm takes is usually an intermediate representation of the program under consideration [28–31]. The first step in slicing a program involves specifying a point of interest, which is called the slicing criterion and is expressed as a tuple $(s, v)$, where $s$ is the statement number and $v$ is the variable that is being used or defined at $s$.

Li et al. [25] proposed the concept of hierarchical slicing that computes the slices at package, class, method, and statement levels. Here, we adopt an approach of slicing that is different from that given in [1, 25]. We name this slicing approach *hierarchical decomposition (HD)* slicing. We first construct a single intermediate graph of the program taking into account the possible dependences among different program elements. Then, we perform HD slicing to obtain the affected program parts with respect to the change made to the program. The slice thus obtained is graphically represented and named *affected slice graph (ASG)*. The steps of HD slicing are given in [32]. A comparison of hierarchical slicing [1] versus HD slicing in terms of number of nodes selected and computation time is shown in Table 2. At each level, we obtain more accuracy in test case selection from a coarse-grain level to a finer-grain level by discarding the test cases that are not relevant.

*2.2. Coupling in Object-Oriented Programming.* Coupling is defined as the degree of interdependence between two modules. However, in an object-oriented programming environment, coupling can exist not only at the level of methods but also at the class level and package level. Therefore, coupling represents the degree of interdependence between methods, between classes, between packages, and so forth. Many researchers have proposed different slicing based mechanisms [19, 20, 32] to measure coupling in an object-oriented framework. There are many factors, such as information hiding, encapsulation, inheritance, message passing,

```
package pkg;
importjava.util.*;
public class TestShape{
  public static void main(String[] args){
    String str;
    int a, b;
    Scanner sin = new Scanner(System.in);
    System.out.println("Enter the Color: ");
    str = sin.next();
    System.out.println("Enter the length and breadth: ");
    a = sin.nextInt();
    b = sin.nextInt();
    Shape s1 = new Rectangle(str, a, b);
    System.out.println(s1);
    System.out.println("Area is " + s1.getArea());
    System.out.println("Enter the Color: ");
    str = sin.next();
    System.out.println("Enter the length and breadth: ");
    a = sin.nextInt();
    b = sin.nextInt();
    Shape s2 = new Triangle(str, a, b);
    System.out.println(s2);
    System.out.println("Area is " + s2.getArea());}}
package pkg;
public class Triangle <T> extends Shape{
private T base;
private T height;
public Triangle(String color, T base, T height){
  super(color);
  this.base = base;
  this.height = height; }
public String toString(){
  return "Triangle of base=" + base + " and height=" + height + ", subclass of " + super.toString(); }
public T getArea(){
  return 0.5*base*height; }}
package pkg;
public class Rectangle extends Shape{
private int length;
private int width;
public Rectangle(String color, int length, int width){
  super(color);
  this.length = length;
  this.width = width; }
public String toString(){
  return "Rectangle of length=" + length + " and width=" + width + ", subclass of " + super.toString(); }
public double getArea(){
  return length*width; }}
package pkg;
public class Shape{
private String color;
public Shape (String color){
  this.color = color; }
public String toString(){
  return "Shape of color=\"" + color + "\""; }
public double getArea(){
  System.err.println("Shape unknown! Cannot compute area!");
  return 0; }}
```

Algorithm 1: An example Java program.

Table 2: Comparison of hierarchical slicing [1] versus HD slicing.

| Sl. number | Program | # nodes | Hierarchical slicing | | HD slicing | |
|---|---|---|---|---|---|---|
| | | | # selected nodes | Time (ms) | # selected nodes | Time (ms) |
| 1 | Expt Prog. | 89 | 35 | 16.6 | 33 | 15.52 |
| 2 | Stack | 157 | 66 | 19.87 | 63 | 18.56 |
| 3 | Sorting | 183 | 77 | 19.98 | 69 | 18.77 |
| 4 | BST | 185 | 74 | 19.98 | 71 | 18.79 |
| 5 | Crc | 331 | 148 | 23.37 | 142 | 21.83 |
| 6 | DLL | 377 | 183 | 23.96 | 171 | 22.16 |
| 7 | ATM | 543 | 241 | 24.81 | 237 | 24.08 |

and abstraction mechanisms, that contribute to coupling in object-oriented programs. High coupling affects program comprehension and analysis. As a result, it becomes very difficult to maintain software systems. In an object-oriented program, coupling can exist between any two components due to message passing, polymorphism, and inheritance mechanisms of object-oriented programs. These components include packages, classes, methods, and statements. Two statements $s1$ and $s2$ are said to be coupled if $s1$ has some dependence (control, data, or type dependence) on $s2$. Methods in an object-oriented program belong to the constituent classes. It implies that a method is coupled either with a method in the same class or with another method in a different class. If the methods of any two classes are coupled, then the corresponding classes are said to be coupled. Similarly, the container packages of the coupled classes are also said to be coupled. The coupling mechanism adopted in this paper is given in Section 3.

*2.3. Regression Test Case Prioritization.* Testing is an important phase in the software life cycle. This phase incurs approximately 60% of the total cost of the software. Therefore, it becomes highly essential to devise proper testing techniques in order to design test cases that tests the software to detect early bugs. It becomes a big challenge to manage the retesting process with respect to the time and cost, especially when the test suite becomes too large. Therefore, selective retest technique attempts to identify those test cases that can exercise the modified parts of the program and the parts that are affected by the modification to reduce the cost of testing. However, test case prioritization can complement selective retest technique and faults can be detected early by prioritizing these selected test cases. Thus, test case prioritization (TCP) problem, stated by Rothermel et al. [6], is as follows, given that $T$ is a test suite; $PT$ is the set of permutations of $T$; $f$ is a function from $PT$ to the real numbers.

*Problem*. Find $T' \in PT$ such that

$$\left(\forall T''\left(T'' \in PT\right) \cap \left(T'' \neq T'\right)\left[f\left(T'\right) \geq \left(T''\right)\right]\right), \quad (1)$$

where $PT$ is the set of all possible orderings of the test cases in $T$ and $f$ is a function that maps the ordering with an award value.

This prioritization approach can be used with the selective retest technique to obtain a version specific prioritized test suite [2]. Rothermel et al. [6] proposed a metric to ensure the efficiency of any of the existing prioritizing techniques. This metric is named as Average Percentage of Fault Detected (APFD) and is used by many researchers to evaluate the effectiveness of their proposed techniques. APFD measure is calculated by taking the weighted average of the number of faults detected during execution of a program with respect to the percentage of test cases executed. Let $T$ be a test suite and let $T'$ be a permutation of $T$. The APFD for $T'$ is defined as follows:

$$\text{APFD} = 1 - \frac{\sum_{i=1}^{n-1} F_i}{n * l} + \frac{1}{2n}. \quad (2)$$

Here, $n$ is the number of test cases in $T$, $l$ is the total number of faults, and $F_i$ is the position of the first test case that reveals the fault $i$. The value of APFD can range from 0 to 100 (in percentage). The higher the APFD value for any ordering of the test cases in the test suite is, the higher the rate at which software faults are discovered is.

## 3. Our Proposed Approach

In this section, we discuss our proposed approach to prioritize a given test suite based on the test cases selected for regression testing. We consider the example Java program shown in Algorithm 1 to discuss our proposed approach. This program defines a class named *Shape* which is inherited by the classes *Rectangle* and *Triangle*. The class *TestShape* contains the main method and computes the area of a rectangle and triangle, respectively, based on the user inputs given through the console, and displays the result. Though this program is very small in size, it represents all the important features of a Java program and is helpful in explaining the working of this approach. The prioritization steps are summarized as follows.

*Step 1.* Construct the ASG and compute the coupling value of each node of the ASG.

*Step 2.* Cluster the coupling values and assign weight to the nodes of ASG.

*Step 3.* Compute the weights of test cases and prioritize.

*3.1. ASG and Computation of Coupling Values.* ASG is the graphical representation of the slice that is computed with respect to some change made to the program. The point of change is taken as the slicing criterion to compute the slice. The slicing algorithm comprises both forward and backward traversal to discover the affected program parts. The forward traversal discovers the program parts affected by the change, and the backward traversal discovers those parts that affect the parts discovered in the forward traversal. The steps of *hierarchical decomposition (HD) slicing* to compute the slice and construct the ASG are given as follows:

(i) Traverse the EOOSDG in forward direction, starting from the point of modification, that is, slicing criterion, except *method overridden* edges.

(ii) Mark and Add each node of the EOOSDG that is reached by the forward traversal to a worklist, $Q_1$.

(iii) Perform two-pass backward traversal for each $q \in Q_1$ as the starting point.

(1) *Pass-1:*

(a) Traverse backward from $q$ through the corresponding edges, avoiding the following edges: *polymorphic call edge, inherited membership edge, parameter_out edge,* and *generic_out edge* to extract all those nodes on which node $q$ depends.

(b) Mark and Add every node $n'$ reached during backward traversal to a worklist, $Q_2$.

(2) *Pass-2:*

(a) Traverse backward from each $q' \in Q_2$ avoiding the following edges *parameter_in edge*, *generic_in edge*,and *any edge traversed in Pass-1.*

(b) Mark and Add every node $n''$ reached during backward traversal to a worklist, $Q_3$.

(iv) Compute final slice as the union of all the marked nodes, $Q = Q_1 \cup Q_2 \cup Q_3$.

(v) To obtain the hierarchical slice, we do the following:

(a) Find $P_1 = P \cap Q$, where $P$ is the set of packages in the program and $P_1$ is the set of affected packages.

(b) Update $Q = Q - P_1$; now $Q$ contains classes, methods, and statements.

(c) Find $C_1 = C \cap Q$, where $C$ is the set of classes in the program and $C_1$ is the affected classes.

(d) Update $Q = Q - C_1$; now $Q$ contains only the methods and statements.

(e) Find $M_1 = M \cap Q$, where $M$ is the set of methods in the program and $M_1$ is the affected methods.

(f) Update $Q = Q - M_1$; now $Q$ contains only affected statements.

(g) Set $S_1 = Q$, where $S_1$ is the set of affected statements.

This set of affected nodes and their dependences are then modeled graphically to form the *affected slice graph (ASG)*. To avoid repetition of the concepts, details are not reproduced here; interested readers are requested to refer to [32] for details.

Algorithm 2 takes the ASG as input and calculates the ACC of each node. We discuss the working of the proposed Algorithm 2 in Section 3.4. In this approach, we use the concept of information inflow and outflow for coupling measurement. The ASG represents all forms of information flow between any two nodes in the form of edges. Thus, our proposed affected component coupling (ACC) for a given node $n$ is computed as the normalized ratio of the sum of inflow and outflow of $n$ with total nodes in ASG. The direction of couplings between any two nodes is given equal weight and was not considered separately. This goes with the hypothesis that a ripple change can propagate in any direction along a coupling dimension. Below, we define the terms related to the computation of affected component coupling (ACC) values.

*Definition 1.* To measure the coupling, we define a set Inflow($n$) that comprises all those nodes on which $n$ depends. For any node $n$ in ASG,

$$\begin{aligned}\text{Inflow}(n) &= \{n_1, n_2, \ldots, n_k \mid \langle n_1, n_2\rangle, \langle n_2, n_3\rangle, \ldots, \langle n_k, n\rangle \\ &\in E_a \land n_1, n_2, \ldots, n_k,\ n \in N_a \land 1 \le k \le |N_a| - 1\}.\end{aligned} \tag{3}$$

The outflow of $n$ in ASG is defined as the set comprising all those nodes that depends on $n$:

$$\begin{aligned}\text{Outflow}(n) &= \{n_1, n_2, \ldots, n_l \mid \langle n, n_1\rangle, \langle n_1, n_2\rangle, \ldots, \langle n_{l-1}, n_l\rangle \\ &\in E_a \land n_1, n_2, \ldots, n_l,\ n \in N_a \land 1 \le l \le |N_a| - 1\}.\end{aligned} \tag{4}$$

Thus, the dependence set $\psi(n)$ of each node is defined as the union of all the Inflow($n$) and Outflow($n$):

$$\psi(n) = \text{Inflow}(n) \cup \text{Outflow}(n). \tag{5}$$

*Definition 2.* Hence, affected coupling of a given node $n$ is computed as the normalized ratio of dependence of $n$, $\psi(n)$, to the total number of affected nodes in the ASG, $|N_a| - 1$, as the node under consideration is excluded. This coupling is measured with respect to the change made to the program that was taken as slicing criterion to generate ASG. This coupling measure is given as

$$\text{ACC}(n) = \frac{|\psi(n)|}{|N_a| - 1}. \tag{6}$$

*Definition 3.* The updated coupling of a method node $M$ in ASG $G_a = (N_a, E_a)$ is defined as the average of the coupling values of all its elements (parameters and statements) along with its own coupling measure. Let a method node $M$ have $j$ number of elements, that is, $n_1, n_2, \ldots, n_j$. Thus, coupling of the method node $M$ is given as

$$\text{ACC}(M) = \frac{\text{ACC}(M) + \sum_{i=1}^{j} \text{ACC}(n_i)}{j + 1}. \tag{7}$$

**Input**: Affected Component Dependency Graph (ASG), total number of nodes $n$
**Output**: Weighted Export Coupling Factor (WACC) of each node
(1) for $x := V_1, V_2, V_3, \ldots, V_n$ (Where $x$ is any node in ASG)
(2) Setstatus$_x$ = FALSE
(3) outflow := call FTraverse(ASG, $x$)
(4) inflow := call BTraverse(ASG, $x$)
(5) $\text{ACC}[x] := \dfrac{(i \text{ nodes} + e \text{ nodes})}{n-1}$
(6) End for (To update the coupling value of all the method, class and package nodes)
(7) for $u := M_1, M_2, M_3, \ldots, M_m$ (Where $m$ is the number of method nodes in the graph)
(8) $\text{ACC}[u] := \dfrac{\left(\text{ACC}[u] + \sum_{i=1}^{j} \text{ACC}[n_i]\right)}{(j+1)}$ ($n_i$ is the statement/parameter node of method $M_i$, $j$ is the total number of statement/parameter nodes of each $M_i$)
(9) End for
(10) for $u := C_1, C_2, C_3, \ldots, C_c$ ($c$ is the total number of class nodes)
(11) $\text{ACC}[u] := \dfrac{\left(\text{ACC}[u] + \sum_{i=1}^{k} \text{ACC}[n_i]\right)}{(k+1)}$ ($n_i$ is the attribute/method node of class $C_i$, $k$ is the total number of attribute/method nodes of each $C_i$)
(12) End for
(13) for $u := P_1, P_2, P_3, \ldots, P_p$ ($p$ is the total number of package nodes)
(14) $\text{ACC}[u] := \dfrac{\left(\text{ACC}[u] + \sum_{i=1}^{l} \text{ACC}[n_i]\right)}{(l+1)}$ ($n_i$ is the subpackage/class node of package $P_i$, $l$ is the total number of subpackage/class nodes of each $P_i$)
(15) End for
(16) $\text{ACC}(S) = \dfrac{\sum_{i=1}^{|N_a|} \text{ACC}(n_i)}{|N_a|}$ (ACC($S$) represents the cohesion of slice $S$) (To assign a weight to each node of ASG)
(17) for $u := V_1, V_2, V_3, \ldots, V_n$ (Where $u$ is any node in ASG)
(18) if ACC[$u$] $\geq 0.7$ and ACC[$u$] $\leq 1.0$
(19) WACC[$u$] := 3
(20) End if
(21) else if ACC[$u$] $\geq 0.6$ and ACC[$u$] $< 0.7$
(22) WACC[$u$] := 2
(23) End else if
(24) else
(25) WACC[$u$] := 1
(26) End else
(27) Exit

Algorithm 2: findWACC(ASG, $n$).

*Definition 4.* The updated coupling of a class node $C$ in ASG $G_a = (N_a, E_a)$ is defined as the average of the coupling values of all its elements (attributes and methods) along with its own coupling measure. Let a class node $C$ have $k$ number of elements, that is, $n_1, n_2, \ldots, n_k$. Thus, cohesion of the class node $C$ is given as

$$\text{ACC}(C) = \frac{\text{ACC}(C) + \sum_{i=1}^{k} \text{ACC}(n_i)}{k+1}. \tag{8}$$

*Definition 5.* The updated coupling of a package node $P$ in ASG $G_a = (N_a, E_a)$ is defined as the average of the coupling values of all its elements (classes and subpackages) along with its own coupling measure. Let a package node $P$ have $l$ number of elements, that is, $n_1, n_2, \ldots, n_l$. Thus, coupling of the package node $P$ is given as

$$\text{ACC}(P) = \frac{\text{ACC}(P) + \sum_{i=1}^{l} \text{ACC}(n_i)}{l+1}. \tag{9}$$

The detail computation of ACC value for some of the nodes is shown in Section 3.4. The reason for taking coupling into consideration is that any node having higher ACC value is an indicator that the node is likely to be more error-prone [20]. This is because higher ACC value of a node indicates more dependence of other nodes on this source of information.

*3.2. Clustering and Assigning Weights.* Once the ACC values of all the nodes have been computed, then the values are clustered. Clustering [33] of the nodes is based on the notion that not all the nodes of ASG are equally erroneous. Some nodes are more erroneous than others. So, we need to identify the set of nodes that can be categorized into different levels of fault-proneness. $k$-means clustering technique [22, 34] is used here to cluster the ACC values. $k$-means clustering is a technique of automatically partitioning a set of given data into $k$ groups. The $k$ cluster centres are chosen randomly

**Input**: Test Suite $T$ with coverage information, Weight of each node in ASG
**Output**: Prioritized Test Suite $T'$
(1) Set $T' = \text{NULL}$
(2) for each test case $t \in T$ do
(3) $\text{Wtc}(t) = \sum_{i=1}^{j} \text{Wt}(V_{ci}(t))$ (Where $V_{ci}(t)$ is the node covered by $t$ whose weight is 3, $\text{Wtc}(t)$ is the total weight of $j$ critical fault prone nodes covered by $t$)
(4) $\text{Wtm}(t) = \sum_{i=1}^{k} \text{Wt}(V_{mi}(t))$ (Where $V_{mi}(t)$ is the node covered by $t$ whose weight is 2, $\text{Wtm}(t)$ is the total weight of $k$ moderate fault prone nodes covered by $t$)
(5) $\text{Wtw}(t) = \sum_{i=1}^{l} \text{Wt}(V_{wi}(t))$ (Where $V_{wi}(t)$ is the node covered by $t$ whose weight is 1, $\text{Wtw}(t)$ is the total weight of $l$ weak fault prone nodes covered by $t$)
(6) $\text{Wt}(t) = \text{Wtc}(t) + \text{Wtm}(t) + \text{Wtw}(t)$ (Where $\text{Wt}(t)$ is the total weight of $t$)
(7) End for (sort on the basis of $\text{Wt}(t)$)
(8) $T' = \text{sort}(T_{\text{Wt}(t)})$
(9) if $\exists t_i, t_j \in T'$ s.t $\text{Wt}(t_i) = \text{Wt}(t_j), i \neq j$ (sort on the basis of $\text{Wtc}(t_i)$, $\text{Wtc}(t_j)$)
(10) $T' = \text{sort}\left(T_{\text{Wtc}(t_i,t_j)}\right)$
(11) if $\text{Wtc}(t_i) = \text{Wtc}(t_j), i \neq j$ (sort on the basis of $\text{Wtm}(t_i)$, $\text{Wtm}(t_j)$)
(12) $T' = \text{sort}\left(T'_{\text{Wtm}(t_i,t_j)}\right)$
(13) if $\text{Wtm}(t_i) = \text{Wtm}(t_j), i \neq j$ (sort on the basis of $\text{Wtw}(t_i)$, $\text{Wtw}(t_j)$)
(14) $T' = \text{sort}\left(T'_{\text{Wtw}(t_i,t_j)}\right)$
(15) Exit

Algorithm 3: H-PTCACC($T$, WACC).

from the data set. The value of $k$ for our approach is 3 as we divide the coupling values into three categories as shown in Figure 3. These three categories of fault association are critical fault association, moderate fault association, and weak fault association. The computed ACC values can belong to either of these three categories. We propose an algorithm named *find Weighted Affected Component Coupling (findWACC)*. It takes the ASG and its total number of nodes as input. It uses the formula given in (6) to compute the ACC value of each node in the ASG. It computes the outflow of a node at Line (3) and inflow of a node at Line (4). Algorithm 2 computes the ACC values of each node and then updates these values for some specific nodes such as package nodes, class nodes, method nodes, and method call nodes. It then assigns weight to the nodes of ASG. Any value of weights can be chosen to signify the faultiness of one set of nodes compared to the other sets. However, in this paper, we use the following weights: if the ACC value of a node $x$ belongs to the category of critical fault association, that is, $0.7 \leqslant \text{ACC}(x) < 1.0$, then $x$ is assigned a weight 3. Similarly, if ACC value of a node $x$ belongs to the category of moderate fault association, that is, $0.6 \leqslant \text{ACC}(x) < 0.7$, then $x$ is assigned a weight 2. Otherwise, $x$ belongs to the category of weak fault association and is assigned a weight 1. The ACC value of each node of ASG and the corresponding weights assigned to them are shown in Figure 2.

*3.3. Computation of Test Case Weights and Prioritization.* The program under consideration is executed with each selected test case in a given test suite to find the coverage information as shown in Table 3. The weight of a test case depends upon the weight of the nodes that it covers. All the critical and moderate nodes (nodes with weights 3 and 2, resp.) are shown in bold in the nodes covered column of Table 3. We propose Algorithm 3 named *Hierarchical Prioritization of Test Cases using Affected Component Coupling (H-PTCACC)* to compute the weights and prioritize the given test suite. Algorithm 3 takes the selected test cases along with their coverage information and ACC values of each node in the ASG as its input. The output of the algorithm is a prioritized set of test cases. For any test case $t_i \in T$, Algorithm 3 first computes its critical weight (Wtc), that is, the sum of the weights of all the critical fault-prone nodes covered by $t_i$. Similarly, Algorithm 3 computes the moderate weight (Wtm), that is, the sum of the weights of all the moderate fault-prone nodes covered by $t_i$. In the same way, Algorithm 3 computes the weak weight (Wtw), that is, the sum of the weights of all weak fault-prone nodes for each test case. Thus the weight of test case is given as the sum of its critical weight (Wtc), moderate weight (Wtm), and weak weight (Wtw). Table 4 shows the different weights computed for each of the test cases $T6$, $T7$, $T8$, $T9$, and $T10$. Algorithm 3 assigns priority to the test cases based on their different computed weights. The test case having a higher total weight is given higher priority in the test suite. If any of the two test cases have the same total weight then their priority is decided based on their critical weight. The test case with higher critical weight is given higher priority. Similarly, if the critical weights are also the same then the moderate weights are taken into consideration for prioritization. If the moderate weight of the test cases is again the same then the weak weights are considered for prioritization. If the weak weights are still the same for any two test cases, then both of the test cases are given equal priority. The last column in Table 4 shows the final case, that is, if the weak weights are still the same for any two test cases, then both of the test cases are given equal priority. The last column in Table 4 shows the final prioritized sequence of the selected test cases.

Table 3: Test case coverage of fault-prone affected nodes.

| Sl. number | Test case | Nodes covered | # nodes | Test case weight |
|---|---|---|---|---|
| 1 | $T6$ | **1, 2, 3, 4, 6, 7** | 6 | 17 |
| 2 | $T7$ | **1, 2, 21, 46, 27, f3, f4, 29, 30, f27_1_out, f27_2_out, 33, 34, f3_out, A3_out**, 24, 25, 26, A5, A6 | 20 | 41 |
| 3 | $T8$ | **1, 2, 3, 4, 6, 7, 21, 46, 27, f3, f4, 29, 30, f27_1_out, f27_2_out**, 19, 20, A5, A6, 25, 26 | 21 | 41 |
| 4 | $T9$ | **1, 2, 3, 4, 6, 7, 21, 46, 52, 27, f4, 33, 30, f3_out, 34**, 24, A6 | 17 | 40 |
| 5 | $T10$ | **1, 2, 3, 4, 21, 23, A3_out, 46, 34, 33, f3_out**, 24 | 12 | 31 |

Table 4: Distribution of test case weights on the basis of fault-prone impact.

| Sl. number | Test case | Weight of critical fault-prone nodes covered | Weight of moderate fault-prone nodes covered | Weight of weak fault-prone nodes covered | Total weight of test case | Priority (according to the total wt. of test case) |
|---|---|---|---|---|---|---|
| 1 | $T6$ | 15 | 2 | 0 | 17 | V |
| 2 | $T7$ | 18 | 18 | 5 | 41 | I |
| 3 | $T8$ | 15 | 20 | 6 | 41 | II |
| 4 | $T9$ | 24 | 14 | 2 | 40 | III |
| 5 | $T10$ | 24 | 6 | 1 | 31 | IV |

*3.4. Working of the Algorithm.* In this subsection, we discuss the working of our proposed algorithms. Algorithm 2 uses the formula given in (6) to compute the ACC value of each node in the ASG. For example, we show the ACC calculation for the class Triangle represented as node 24 in Figure 1. Initially, ACC value of node 24 is computed as

$$\mathrm{ACC}(24) = \frac{|\mathrm{outflow}(24)| + |\mathrm{inflow}(24)|}{|N_a| - 1} = \frac{20+4}{32} = 0.75. \quad (10)$$

Figure 4 shows the sets associated with the computation of ACC of node 24. The inflow set for node 24 is shown in Figure 4(a) and outflow set is shown in Figure 4(b). Figure 4(c) shows the set in which node 24 is a member. Once the computation of ACC of all member nodes of node 24, shown in Figure 4(d), is complete then ACC(24) is updated. Similarly, the ACC values of all the associated nodes (25, 26, 27, f3, f4, 29, 30, f27_1_out, f27_2_out, 33, f3_out, and 34) with node 24 as shown in Figure 1 are computed as follows:

$$\mathrm{ACC}(25) = \frac{|\mathrm{outflow}(25)| + |\mathrm{inflow}(25)|}{|N_a| - 1} = \frac{3+14}{32} = 0.5312,$$

$$\mathrm{ACC}(26) = \frac{|\mathrm{outflow}(26)| + |\mathrm{inflow}(26)|}{|N_a| - 1} = \frac{3+14}{32} = 0.5312,$$

$$\mathrm{ACC}(27) = \frac{|\mathrm{outflow}(27)| + |\mathrm{inflow}(27)|}{|N_a| - 1} = \frac{12+7}{32} = 0.5937,$$

$$\mathrm{ACC}(f3) = \frac{|\mathrm{outflow}(f3)| + |\mathrm{inflow}(f3)|}{|N_a| - 1} = \frac{13+12}{32} = 0.7812,$$

$$\mathrm{ACC}(f4) = \frac{|\mathrm{outflow}(f4)| + |\mathrm{inflow}(f4)|}{|N_a| - 1} = \frac{13+12}{32} = 0.7812,$$

$$\mathrm{ACC}(29) = \frac{|\mathrm{outflow}(29)| + |\mathrm{inflow}(29)|}{|N_a| - 1} = \frac{12+13}{32} = 0.7812,$$

$$\mathrm{ACC}(30) = \frac{|\mathrm{outflow}(30)| + |\mathrm{inflow}(30)|}{|N_a| - 1} = \frac{12+13}{32} = 0.7812,$$

$$\mathrm{ACC}(f27_1_out) = \frac{|\mathrm{outflow}(f27_1_out)| + |\mathrm{inflow}(f27_1_out)|}{|N_a| - 1} = \frac{10+14}{32} = 0.75,$$

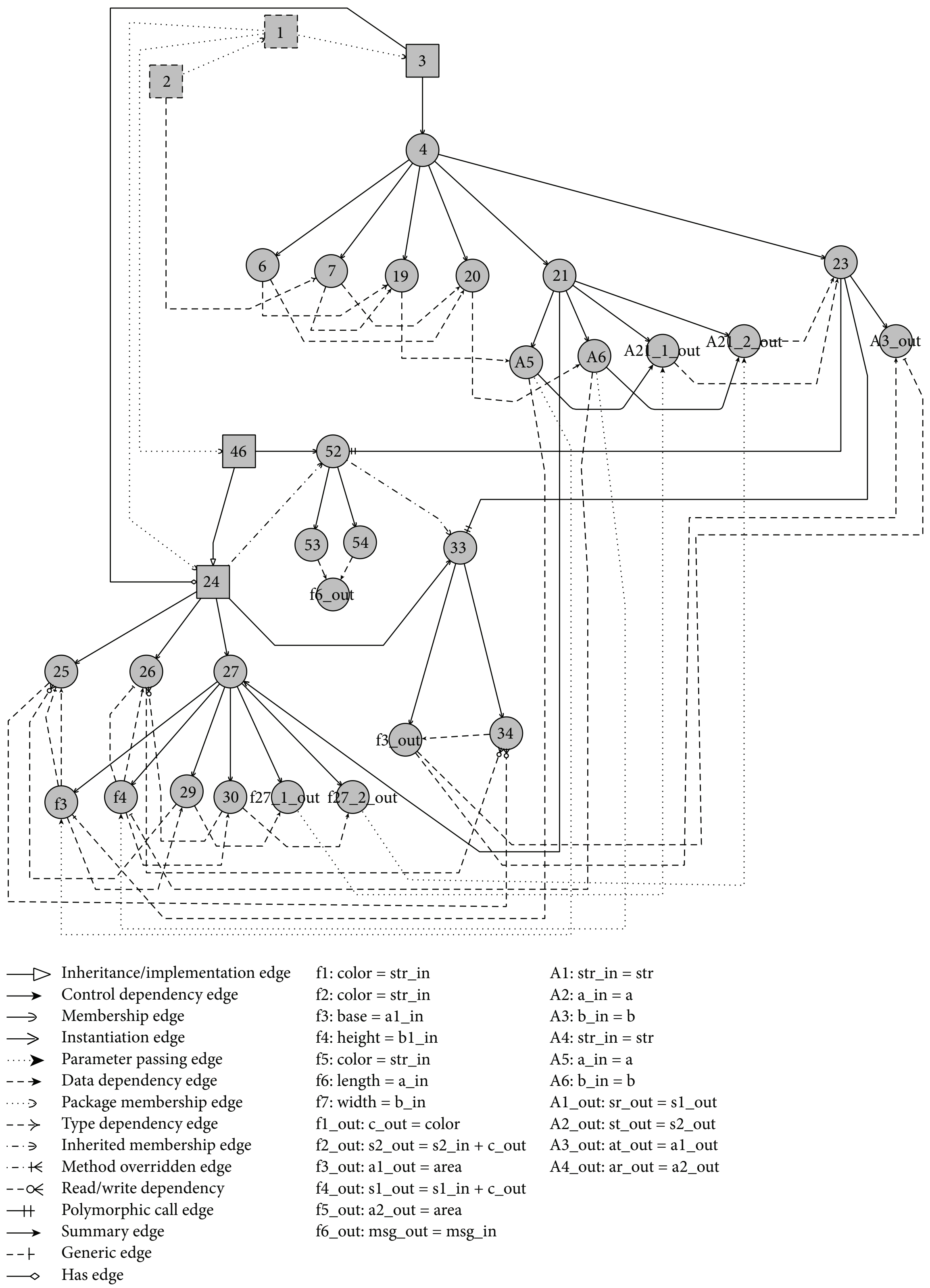

Figure 1: Affected slice graph (ASG) of the example Java program given in Algorithm 1.

$$\begin{aligned}
&\text{ACC (f27_2_out)} \\
&= \frac{|\text{outflow (f27_2_out)}| + |\text{inflow (f27_2_out)}|}{|N_a| - 1} \\
&= \frac{10 + 14}{32} = 0.75, \\
&\text{ACC (33)} = \frac{|\text{outflow (33)}| + |\text{inflow (33)}|}{|N_a| - 1} = \frac{3 + 24}{32} \\
&= 0.8437, \\
&\text{ACC (f3_out)} \\
&= \frac{|\text{outflow (f3_out)}| + |\text{inflow (f3_out)}|}{|N_a| - 1} = \frac{1 + 28}{32} \\
&= 0.9062, \\
&\text{ACC (34)} = \frac{|\text{outflow (34)}| + |\text{inflow (34)}|}{|N_a| - 1} = \frac{2 + 27}{32} \\
&= 0.9062.
\end{aligned} \tag{11}$$

Then, Algorithm 2 updates the ACC value of each node of ASG. The reason behind this update is that, for any node that represents a method, the statements contained inside that method also contribute to the ACC of the method. Even if a method does not have any statement inside it, still it will have some ACC value as some other method may be overriding it. Therefore, we have taken the average of all the ACC values of all the statements and the ACC value of the method under consideration, to compute the updated ACC value of the method. For example, the ACC values of nodes $\{24, 27, 33\}$ are updated. The average ACC value of node 27 along with the ACC values of all its member nodes $\{\text{f3}, \text{f4}, 29, 30, \text{f27_1_out}, \text{f27_2_out}\}$ are computed and assigned to node 27; that is,

$$\begin{aligned}
&\text{ACC (27)} \\
&= \frac{\text{ACC (27)} + \text{ACC (f3)} + \text{ACC (f4)} + \text{ACC (29)}}{7} \\
&+ \frac{\text{ACC (30)} + \text{ACC (f27_1_out)} + \text{ACC (f27_2_out)}}{7} \\
&= \frac{0.5937 + 0.7812 + 0.7812 + 0.7812}{7} \\
&+ \frac{0.7812 + 0.75 + 0.75}{7} = 0.7455.
\end{aligned} \tag{12}$$

Similarly, ACC values of node 33 and node 24 are updated as follows:

$$\begin{aligned}
&\text{ACC (33)} = \frac{\text{ACC (33)} + \text{ACC (f3_out)} + \text{ACC (34)}}{3} \\
&= \frac{0.84375 + 0.90625 + 0.90625}{3} = 0.88542,
\end{aligned}$$

Nodes details: node

| Sr. No. | Node Name | ACC | Weight |
|---|---|---|---|
| 1 | 1 | 0.859375 | 3 |
| 2 | 2 | 0.9296875 | 3 |
| 3 | 3 | 0.953125 | 3 |
| 4 | 4 | 0.8125 | 2 |
| 5 | 6 | 0.84375 | 3 |
| 6 | 7 | 0.84375 | 3 |
| 7 | 19 | 0.65625 | 1 |
| 8 | 20 | 0.65625 | 1 |
| 9 | 21 | 0.7375 | 2 |
| 10 | 23 | 0.921875 | 3 |
| 11 | 24 | 0.69375 | 1 |
| 12 | 25 | 0.53125 | 1 |
| 13 | 26 | 0.53125 | 1 |
| 14 | 27 | 0.77678573 | 2 |
| 15 | 29 | 0.78125 | 2 |
| 16 | 30 | 0.78125 | 2 |
| 17 | 33 | 0.8854167 | 3 |
| 18 | 34 | 0.90625 | 3 |
| 19 | 46 | 0.7833333 | 2 |
| 20 | 52 | 0.8333333 | 2 |
| 21 | 53 | 0.78125 | 2 |
| 22 | 54 | 0.78125 | 2 |
| 23 | A21_1_out | 0.75 | 2 |
| 24 | A21_2_out | 0.75 | 2 |
| 25 | A3_out | 0.90625 | 3 |
| 26 | A5 | 0.6875 | 1 |
| 27 | A6 | 0.6875 | 1 |
| 28 | f27_1_out | 0.75 | 2 |
| 29 | f27_2_out | 0.75 | 2 |
| 30 | f3 | 0.78125 | 2 |
| 31 | f3_out | 0.90625 | 3 |
| 32 | f4 | 0.78125 | 2 |
| 33 | f6_out | 0.8125 | 2 |

FIGURE 2: The calculated ACC values of different nodes of the ASG in Figure 1 and their weights.

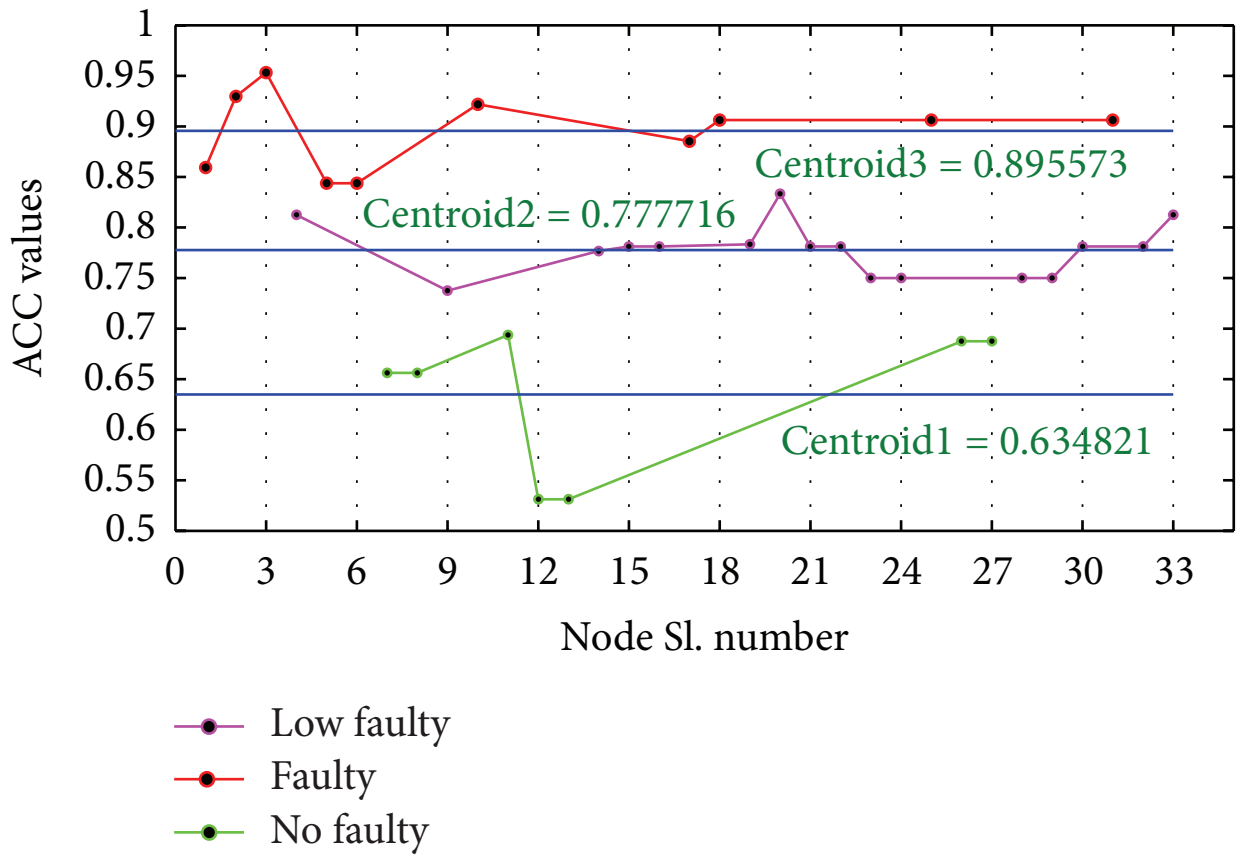

FIGURE 3: $k$-means clustering of the ACC values of the nodes.

$$\begin{aligned}
&\text{ACC (24)} \\
&= \frac{\text{ACC (24)} + \text{ACC (25)} + \text{ACC (26)} + \text{ACC (27)}}{5} \\
&+ \frac{\text{ACC (33)}}{5} \\
&= \frac{0.75 + 0.5312 + 0.5312 + 0.7455 + 0.88542}{5} \\
&= 0.688664.
\end{aligned} \tag{13}$$

Therefore, ACC value of *class Triangle* in Algorithm 1 represented as node 24 in Figure 1 is found to be 0.68866. Similar procedure is followed to update the ACC values of all the nodes representing the classes and packages in the ASG.

Algorithm 3 computes the critical fault-prone weight $\text{Wtc}(t_i)$, moderate fault-prone weight $\text{Wtm}(t_i)$, weak fault-prone weight $\text{Wtw}(t_i)$, and the total weight $\text{Wt}(t_i)$ for each test case $t_i \in T$. For example, the nodes covered by test case $T8$ as given in the second column of Table 3 are {1, 2, 3, 4, 6, 7, 21, 46, 27, f3, f4, 29, 30, f27_1_out, f27_2_out, 19, 20, A5, A6, 25, 26}. The critical fault-prone nodes covered by $T8$ are {1, 2, 3, 6, 7}. So, critical fault-prone weight of $T8$ is calculated as $\text{Wtc}(T8) = \text{Wt}(1) + \text{Wt}(2) + \text{Wt}(3) + \text{Wt}(6) + \text{Wt}(7) = 3+3+3+3+3 = 15$. The moderate fault-prone nodes covered by $T8$ are {4, 21, 46, 27, f3, f4, 29, 30, f27_1_out, f27_2_out}. So, moderate fault-prone weight of $T8$ is calculated as $\text{Wtm}(T8) = \text{Wt}(4) + \text{Wt}(21) + \text{Wt}(46) + \text{Wt}(27) + \text{Wt}(\text{f3}) + \text{Wt}(\text{f4})+\text{Wt}(29)+\text{Wt}(30) + \text{Wt}(\text{f27_1_out})+\text{Wt}(\text{f27_1_out}) = 2 + 2 + 2 + 2 + 2 + 2 + 2 + 2 + 2 + 2 = 20$. Similarly, the weak fault-prone nodes covered by $T8$ are {25, 26, 29, 30} and the weak fault-prone weight of $T8$ is calculated as $\text{Wtw}(T8) = \text{Wt}(19) + \text{Wt}(20) + \text{Wt}(\text{A5}) + \text{Wt}(\text{A6}) + \text{Wt}(25) + \text{Wt}(26) = 1 + 1 + 1 + 1 + 1 + 1 = 6$. Therefore, total weight of the test case $T8$ is calculated as $\text{Wt}(T8) = \text{Wtc}(T8) + \text{Wtm}(T8) + \text{Wtw}(T8) = 15 + 20 + 6 = 41$.

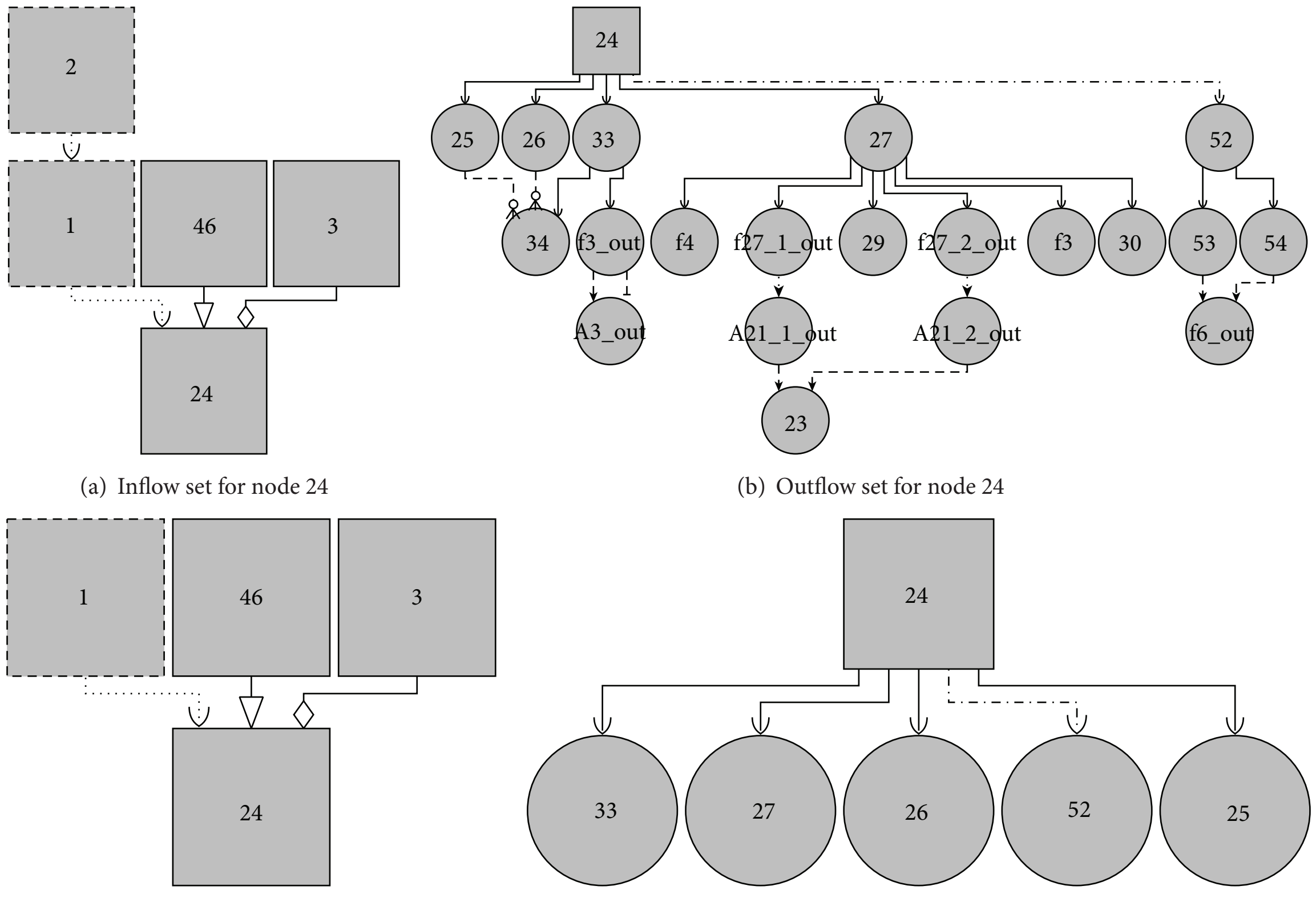

(a) Inflow set for node 24

(b) Outflow set for node 24

(c) The set to which node 24 is a member

(d) The members of node 24

Figure 4: ACC computation of nodes of ASG in Figure 1.

Then, the algorithm sorts the test cases in the decreasing order of their total weights $\text{Wt}(t_i)$. If there exist some test cases $t_i, t_j$ such that $\text{Wt}(t_i) = \text{Wt}(t_j)$, then the algorithm sorts $t_i$ and $t_j$ based on their critical fault-prone weights $\text{Wtc}(t_i)$ and $\text{Wtc}(t_j)$. If for some test cases $\text{Wtc}(t_i) = \text{Wtc}(t_j)$, then $t_i$ and $t_j$ are sorted based on their moderate fault-prone weights $\text{Wtm}(t_i)$ and $\text{Wtm}(t_j)$. If, again, $\text{Wtm}(t_i) = \text{Wtm}(t_j)$, then test cases are sorted by their weak fault-prone weights $\text{Wtw}(t_i)$ and $\text{Wtw}(t_j)$. In a very unlikely case, if the weak fault-prone weights are still identical, that is, $\text{Wtw}(t_i) = \text{Wtw}(t_j)$, then the test cases are given equal priority. The prioritized order of the test cases $T6$–$T10$ based on their respective weights is obtained as $\{T7, T8, T9, T10, T6\}$.

*3.5. Complexity Analysis of the Algorithms.* The complexity analysis of the proposed algorithms is given as follows.

*3.5.1. Space Complexity.* Let the computed slice represented as ASG have $n$ nodes. Each node in the ASG corresponds to each statement of the computed slice along with the actual and formal arguments present. Hence, the space requirement is given as $O(n)$. Each node may have dependences on other nodes. These dependences on other nodes are represented as edges. Since each node can be dependent on maximum $(n-1)$ other nodes, the space requirement for the edges is $O(n^2)$. Hence, the total space requirement for the algorithm is $O(n^2 + n) \equiv O(n^2)$.

*3.5.2. Time Complexity.* Let $n$ be the set of nodes in the ASG. To compute the inflow to the input node, each node is traversed only once, so the time complexity is $O(n)$. If the time spent in each recursive call is ignored, then each node $u$ can be processed in $O(1 + \text{pred}[u])$, where $\text{pred}[u]$ represents the set of predecessor nodes of $u$. If each node has every other node in the graph as its predecessor node, then each node has $(n-1)$ predecessor nodes. So, the time complexity to process each node is $O(1 + (n-1)) \approx O(n)$. Similarly, to compute the outflow from the input node the time complexity is calculated as $O(n)$. Then, the total time required to compute the coupling values of all the nodes is calculated as $O(N^2)$.

Let $m$, $c$, and $p$ be the number of method nodes, class nodes, and package nodes, respectively, whose ACC values need to be updated. If each method node has $j$ member nodes, then the time required to update $m$ method nodes is $O(mjn^2)$. Since $m$ and $j$ are small bounded positive integers, the time complexity is calculated as $O(n^2)$. Similarly, if each class node has $k$ member nodes and each package node has $l$ member nodes, then the respective time complexities for $c$ class nodes and $p$ package nodes are $O(ckn^2)$ and $O(pln^2)$. Since $c$, $k$, $p$, and $l$ are small bounded positive integers, the time complexities are calculated as $O(n^2)$ and $O(n^2)$, respectively, for the class and package nodes. As $n$ nodes are there with $n$ ACC values, so the time required to assign a weight to each of the $n$ nodes depending on their respective ACC value is $O(n)$. Therefore, the worst-case run-time of the findWACC algorithm is calculated as $O(n^2+n^2+n^2+n^2+n) \equiv O(n^2)$.

Let $t$ be the number of test cases to be prioritized in the given test suite $T$. Suppose a test case covers at most $n$ number of nodes. Let $j$, $k$, and $l$ be the critical, moderate, and weak fault-prone nodes, respectively, covered by a test case, such that $n = j + k + l$. So, the time complexity to compute the weight of each test case is calculated as $O(j + k + l) \equiv O(n)$. As a result, the total time complexity to compute the weight of $t$ test cases in the given test suite $T$ is $O(tn)$. Assuming $t \equiv n$, the time complexity to compute the weights is calculated as $O(n^2)$. The time complexity to sort the $t \equiv n$ test cases is calculated as $O(n^2)$. Therefore, the worst-case run-time of the H-PTCACC algorithm is calculated as $O(n^2 + n^2) \equiv O(n^2)$.

## 4. Implementation

In this section, we briefly describe the implementation of our work. We implemented our code and all the algorithms using Java and Eclipse v3.4 IDE on a standard Windows 7 desktop. The proposed approach of change impact analysis is completely based on the intermediate graph of the modified program. The identification of the dependences to construct the intermediate graph follows the build-on-build approach; that is, we use the existing APIs and tools to build the graph instead of developing the source code parser from scratch. Source code instrumentation and generation of the intermediate graph are implemented by using XPath parser on srcML (SouRce Code Markup Language) representation of the input Java program. Thus, srcML is the XML (eXtended Markup Language) representation of the input Java program. The input program is converted to srcML using src2srcml tool. This srcML representation is then used to extract the dependences between program parts by using the XPath parser. The details of the program conversion and fact extraction process can be referred to in [26, 35]. Many other APIs and tools (such as Document Object Model (DOM) and Simple API for XML (SAX)) can be used to extract facts from the srcML representation. In this paper, the fact and dependence extraction is done using XPath. XPath is a language support used by XSLT (extensible stylesheet language) parser [36] to address specific part(s) of the entire XML document. The choice of using XPath is because of its simplicity and easy extraction by direct tracing to the location of the information. This also works on both visioXML and srcML formats of XML. The XPath expression "// function [name = "getArea"]," directly traces to the function definition with the name "getArea." The source code is first instrumented and then dependences in the program are identified and extracted into the program dictionary to construct the intermediate graph. The modified statement (instrumented number) is taken as input along with the intermediate graph, to slice the affected nodes. Most of the dependences at package level, class level, and method level are extracted from the Imagix4D XML data. Imagix4D is a static analysis tool that gives the graphical representation of most of these dependences. The statement level dependences such as control dependence and data dependence [35] are extracted from the srcML representation of the program. The program dictionary stores the following information:

(i) Set of all packages in the program.

(ii) Set of all classes in the program.

(iii) Set of all methods in the program.

(iv) Set of all statements in the program.

(v) Sets of each dependence type.

A change set is maintained that refers to the set of concurrent changes carried out on the program. $k$-means algorithm is implemented in Matlab for clustering the coupling value.

*4.1. Experimental Program Structure.* To implement our technique and show its effectiveness, we have taken total fifteen programs of different specifications as shown in Table 5. Out of these fifteen programs, ten benchmark programs (*Stack*, *Sorting*, *BST*, *CrC*, *DLL*, *Elevator_spl*, *Email_spl*, *GPL_spl*, *Jtopas*, and *Nanoxml*) are taken from Software-Artifact Infrastructure Repository (SIR) [37] and other five programs are developed as academic assignments. These smaller programs are chosen to ascertain the correctness and accuracy of the approach, keeping in mind that they represent a variety of Java features and applications, the test cases are available and otherwise easily developed, and coverage information can be computed.

The smallest program has 54 LOC, and the largest program has 7646 LOC. The total LOC for all the fifteen programs is 19369 and the average LOC per program are 1291. The total number of classes in all the fifteen programs is 185 with an average of 12 classes per program. Our example program in Algorithm 1 has smallest number of classes and *GPL_spl* has the highest, 111 number of classes. The total number of methods in all the programs is 2048 with an average of 137 methods per program. We have constructed a total of 150 ASGs for all the programs. The smallest ASG has 33 nodes, and the largest has 5233 nodes. The total number of affected nodes in all the fifteen programs is 28452, and the average number of nodes affected per each change made to the programs is 152.

Table 5: Result obtained for regression testing of different programs.

| Sl. number | Programs | Lines of code | # classes | # methods | Total # test cases | # mutants | # selected test cases for regression testing | Time for prioritization (sec) |
|---|---|---|---|---|---|---|---|---|
| 1 | Expt. Program | 54 | 4 | 10 | 20 | 14 | 5 | 1.3 |
| 2 | Calculator | 75 | 4 | 24 | 15 | 42 | 7 | 1.8 |
| 3 | Elevator | 90 | 6 | 64 | 25 | 27 | 10 | 2.59 |
| 4 | Stack | 114 | 5 | 24 | 22 | 35 | 9 | 2.27 |
| 5 | Sorting | 130 | 1 | 15 | 16 | 43 | 5 | 1.65 |
| 6 | BST | 130 | 4 | 23 | 20 | 51 | 12 | 3.21 |
| 7 | CrC | 261 | 4 | 34 | 18 | 46 | 6 | 1.64 |
| 8 | DLL | 277 | 1 | 32 | 24 | 47 | 6 | 1.78 |
| 9 | Notepad | 300 | 12 | 68 | 17 | 17 | 8 | 2.07 |
| 10 | ATM | 900 | 24 | 321 | 33 | 39 | 12 | 3.87 |
| 11 | Elevator_spl | 1046 | 17 | 51 | 15 | 53 | 10 | 2.63 |
| 12 | Email_spl | 1233 | 17 | 68 | 18 | 18 | 11 | 2.89 |
| 13 | GPL_spl | 1713 | 111 | 432 | 22 | 22 | 14 | 3.7 |
| 14 | Jtopas | 5400 | 50 | 748 | 16 | 28 | 9 | 2.36 |
| 15 | Nanoxml | 7646 | 24 | 134 | 14 | 32 | 7 | 1.72 |

Similarly, the total number of test cases considered for all the programs is 295 with a mean of 20 test cases per program. Only those test cases that had a coverage value of more than 90% were chosen for each of the experimental programs. The coverage of the test cases were found using JaBUTi, a coverage analysis tool for Java programs. The total number of test cases selected for prioritization using our approach for all the fifteen programs is 131. The smallest number of selected test cases for prioritization is 5 for the our example program in Algorithm 1 and the highest is 14 for *GPL_spl* program.

*4.2. Mutation Analysis.* To generate the mutants for the input program, we used an Eclipse plugin of MuJava known as MuClipse [38]. Fault mutants are considered to be good representative of real faults [37, 39, 40]. MuClipse supports both the traditional and object-oriented operators for mutation analysis. Table 6 gives an overview of the mutation operators considered in the experimental study. A brief description of the operators is given for every operator in Table 6. The first five operators are the traditional operators. The remaining 23 operators relate to the faults in object-oriented programs. Out of which *JTD*, *JSC*, *JID*, and *JDC* are specific to Java features that are not available in all object-oriented languages. Apart from this, there are some other operators, such as *EOA*, *EOC*, *EAM*, and *EMM*, that reflect the typical coding mistakes common during development of an object-oriented software. The mutant generator generates the mutants for the sliced program (representing the affected program parts) according to the operators selected by the testers. Very large number of mutants are generated. The location of these mutants in the source code is visualized through mutant viewer. It allows a tester to select appropriate number of mutants and design test cases to kill the mutants. As the number of generated mutants are too large, we randomly selected a less number of mutants for our experimental programs. This process was repeated for 10 times and the rate of fault detection for the prioritized test suite was computed. The average number of mutants selected for every program is shown in Table 5. The test cases are written in a specific format such that each test case is in a form of invoking a method in the class under test. The test method has no parameters and returns the result in the form of a string. The mutant is said to be killed if the obtained output does not match the output of the original program. The test cases for the input program are generated using JUnit Eclipse plugin as the JUnit test cases closely match the required format. The total number of fault mutants for all the fifteen programs is 514, and the average number of mutants per program is 34.

*4.3. Results.* Figure 5 shows the boxplots of the results of our mutation analysis for all the experimental programs. Figure 5(a) shows the presence of mutants in percentage in the affected parts of the programs. The presence of mutants in the affected parts of the programs ranges from a minimum of 12% (*DLL* program) to a maximum of 94% (*Sorting* program). The affected program parts in five programs have more than 90% of the mutants and four programs have little more than 10% mutants. The result shows that an average of 47% of mutants are scattered in the affected program parts of the sample programs. Figure 5(b) shows the percentage of mutants killed in each of the experimental programs. The percentage of mutants killed by the prioritized test cases varies from 70% to 95%. The average percentage of mutants killed by the prioritized test suite is 85%. This shows that our prioritized test cases are efficient in revealing the faults.

The average percentage of affected nodes covered by the prioritized test cases using the approach of Panigrahi and Mall and our approach is shown in Figures 6 and 7, respectively, for the experimental program given in Algorithm 1. From Figures 6 and 7, it may be observed that the average

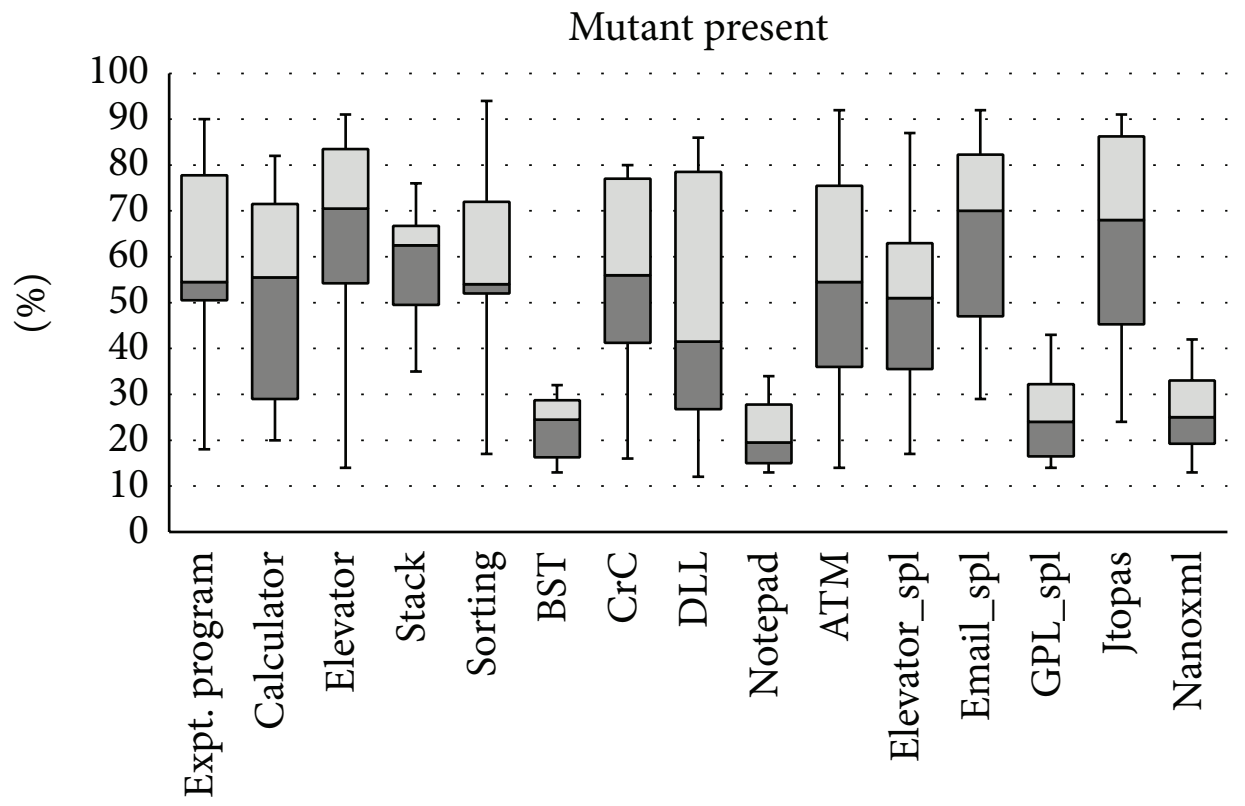

(a) Box-plot of the % of fault mutants present in affected parts of the programs

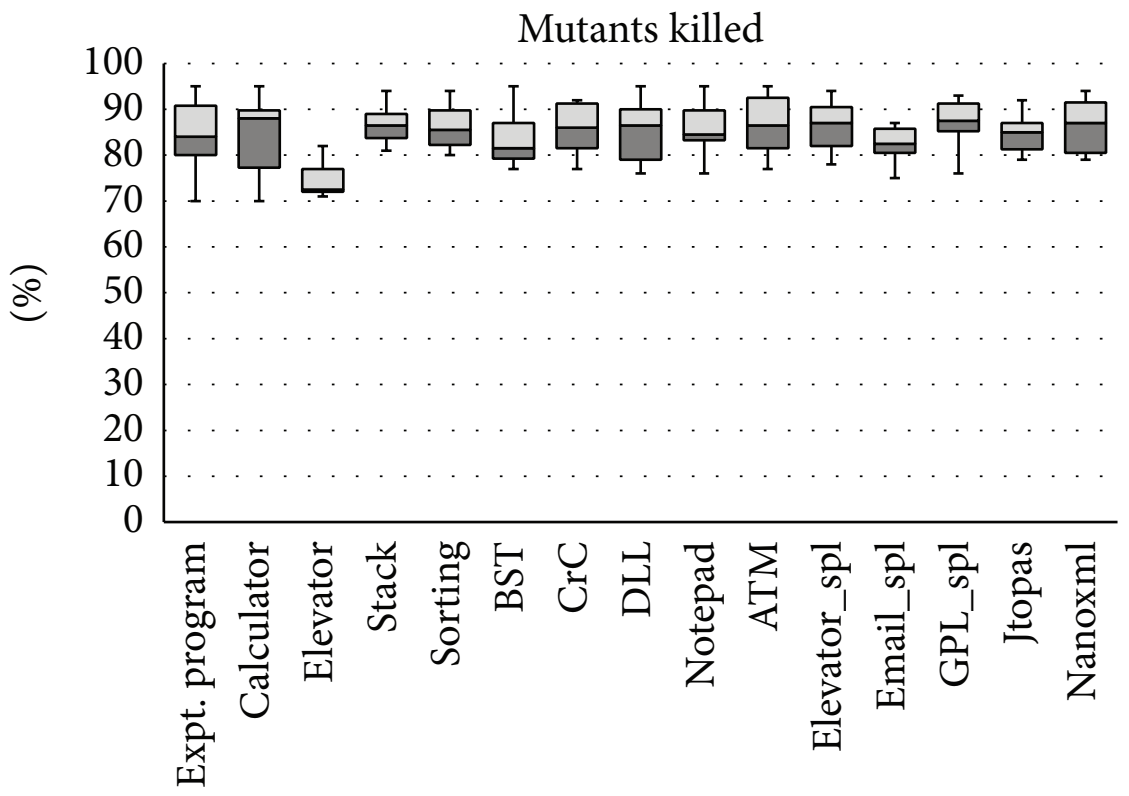

(b) Box-plot of the % of mutants killed in affected parts of the programs

FIGURE 5: Mutation analysis of programs.

TABLE 6: Overview of mutation operators.

| Operator | Description |
|---|---|
| | Traditional operators |
| ABS | Absolute value insertion |
| AOR | Arithmetic operator replacement |
| LCR | Logical connector replacement |
| ROR | Relational operator replacement |
| UOI | Unary operator insertion |
| | Java Interclass operators |
| IHD | Hiding variable deletion |
| IHI | Hiding variable insertion |
| IOD | Overriding method deletion |
| IOP | Overridden method calling position change |
| IOR | Overridden method rename |
| ISK | Super keyword deletion |
| IPC | Explicit call of a parent's constructor deletion |
| PNC | New method call with child class type |
| PMD | Instance variable declaration with parent class type |
| PPD | Parameter variable declaration with child class type |
| PRV | Reference assignment with other compatible types |
| OMR | Overloading method contents change |
| OMD | Overloading method deletion |
| OAO | Argument order change |
| OAN | Argument number change |
| JTD | This keyword deletion |
| JSC | Static modifier change |
| JID | Member variable initialization deletion |
| JDC | Java-supported default constructor creation |
| EOA | Reference and content assignment replacement |
| EOC | Reference and content comparison replacement |
| EAM | Accessor method change |
| EMM | Modifier method change |

percentage of nodes covered (APNC) using the approach of Panigrahi and Mall [18] is 77.2%, whereas the APNC value using our approach is 80.6%. Thus, there is an increase of 3.4% in APNC measure by our approach. Hence, our approach detects faults better than the approach of Panigrahi and Mall [18] as our approach covers more number of fault-prone nodes. We evaluated the effectiveness of our approach by using APFD metric. We named Panigrahi and Mall approach [18] as Affected Node Coverage (ANC) and our approach as Fault-Prone Affected Node Coverage (FPANC) in Figure 8. The comparison of APFD values for these fifteen different programs obtained using ANC and FPANC approaches is shown in Figure 8. The results show that our FPANC approach achieves approximately 8% increase in the APFD metric value over ANC approach.

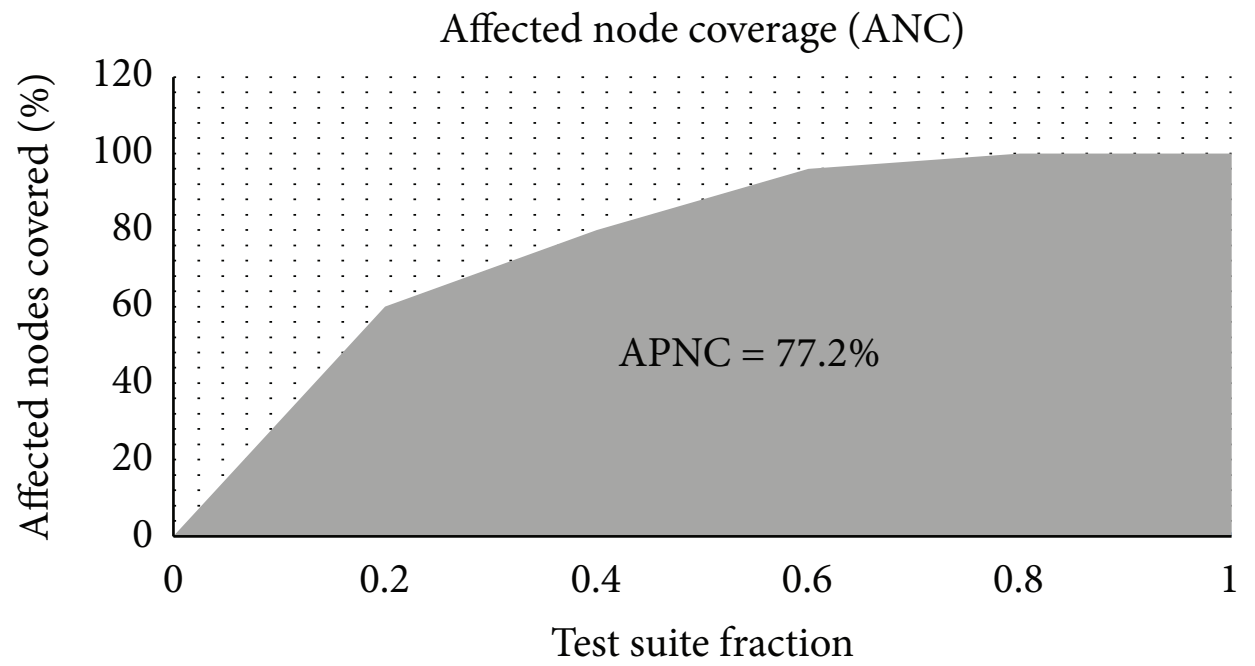

FIGURE 6: Average percentage of affected nodes covered by the prioritized test cases using the approach of Panigrahi and Mall [18].

The experimental results show that the performance of our approach varies significantly with program attributes, change attributes, test suite characteristics, and their interaction. To assume that a higher APFD implies a better technique, independent of cost factors, is an oversimplification that may lead to inaccurate choices among prioritization techniques. For a given testing scenario, cost models for prioritization can be used to determine the amount of difference in APFD that may yield desirable practical benefits, by associating APFD differences with measurable attributes such as prioritization time. A prioritization technique would be acceptable provided the time taken is within acceptable limits, which also reflects the cost of retesting. Korel et al. [41] have also focused on less time of execution to decrease the overhead of prioritization process. However, the acceptable time limit greatly depends upon the testing time available with the tester. An empirical analysis on the prioritization time is outside the scope of this paper and is kept for our future work. We have reported the prioritization time of our approach to indicate the time taken to prioritize the test cases when the precomputed test coverage information and the ASG are available with the tester. The last column of Table 5 shows the time taken for prioritizing the selected test cases. The prioritization time varies from a minimum of 1.3 seconds to a maximum of 3.87 seconds for the experimental programs. The total time taken to prioritize the test cases of all the programs is 35.48 seconds and the average time for prioritizing the test cases is 2.4 seconds. The prioritization time includes the time for computing the weights of the test cases and the time taken to order the test cases in decreasing order of their weights.

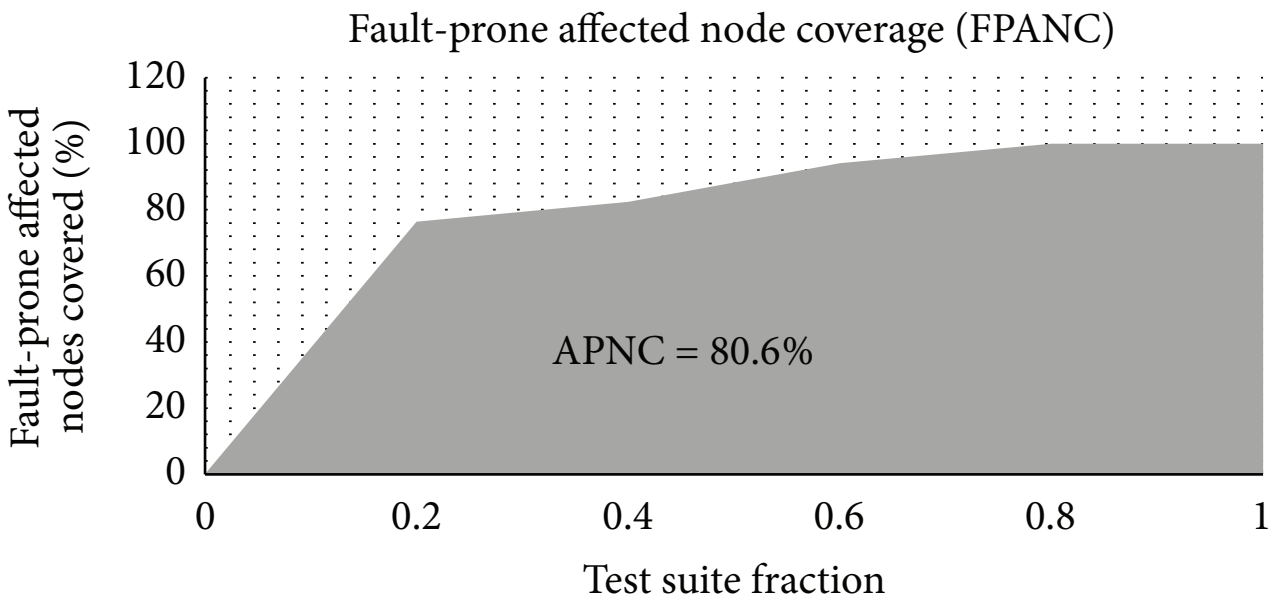

Figure 7: Average percentage of fault-prone affected nodes covered by the prioritized test cases using our approach.

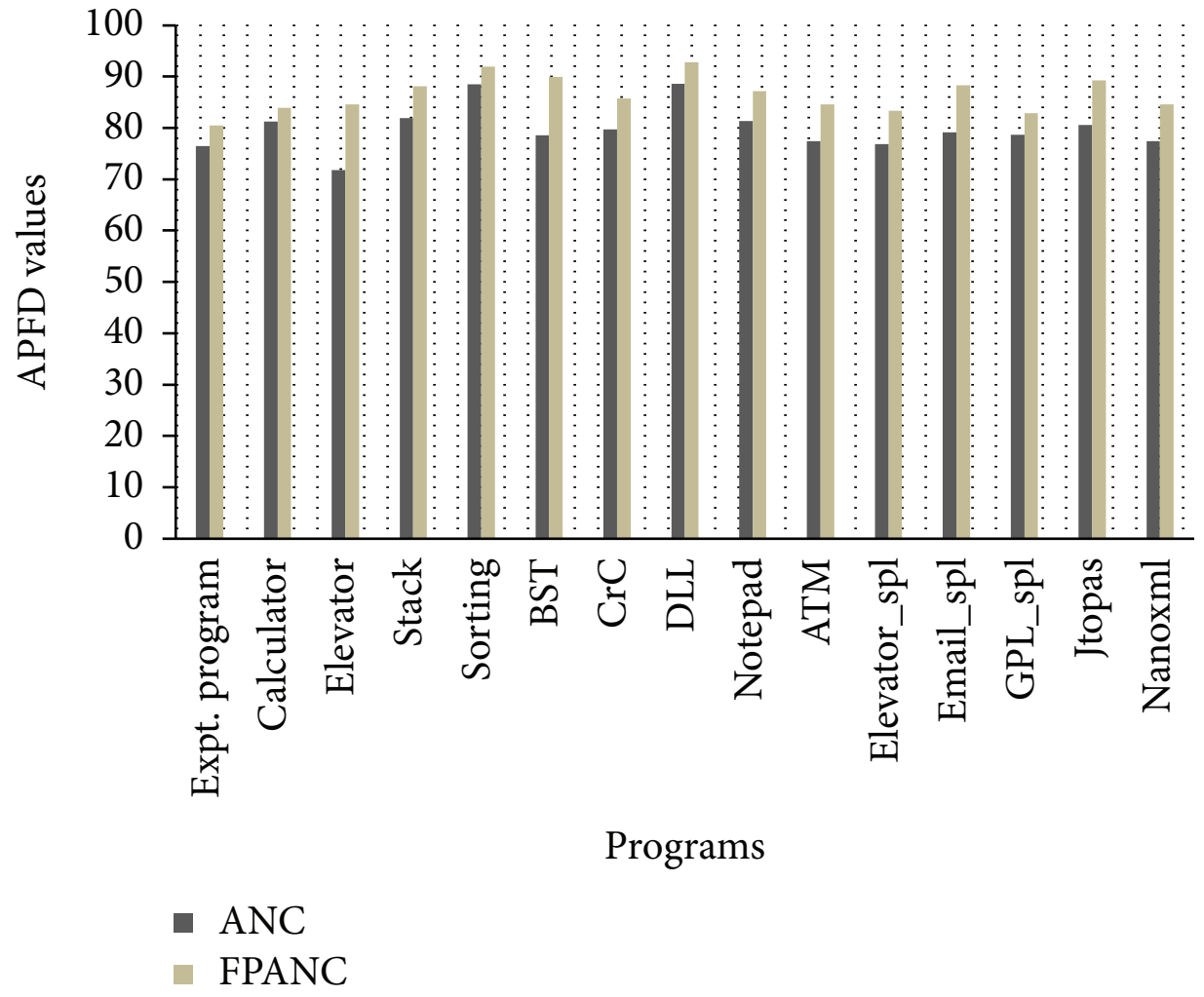

Figure 8: Comparison of APFD values for different programs.

## 5. Comparison with Related Work

In this section, we give a comparative analysis of our work with some other related works.

Elbaum et al. [42] performed an empirical investigation to find out the testing scenarios where a particular prioritization approach will prove to be efficient. They analyzed the rate of fault detection that resulted from several prioritization techniques such as total function coverage, additional function coverage, total binary-diff function coverage, and additional binary-diff function coverage. The authors considered eight C programs for their experimentation. They used the documentation on the programs and the parameters and special effects that they determined to construct a suite of test cases that exercise each parameter, special effect, and erroneous condition affecting program behavior. Then they augmented those test suites with additional test cases to increase code coverage at the statement level. The regression fault analysis was done on the faults inserted by the graduate and undergraduate students with more than two years of coding experience. The experimental results show that the performance of test case prioritization techniques varies significantly with program attributes, change attributes, test suite characteristics, and their interaction. Our results also confirm similar findings. However, our approach concerns Java programs. We have considered the dependencies caused by the object-oriented features in our proposed intermediate graph. Our approach targets the coverage of those affected nodes that have a high potential of exposing faults; hence, it is more change-based than the approach in [42].

Korel et al. [41] proposed a model based test prioritization approach. The approach is based on the assumption that execution of the model is inexpensive as compared to execution of the system; therefore the overhead associated with test prioritization is relatively small. This approach is based on the EFSM system models. The original EFSM model is compared with the modified EFSM model to identify the changes. After the changes are identified, the EFSM model is executed with the test cases to collect different information that are used for prioritization. The authors propose two types of prioritization: selective test prioritization and model dependence-based test prioritization. The selective test prioritization approach assigns higher priority to the test cases that execute the modified transitions. Model dependence-based test prioritization mechanism carries out dependency analysis between the modified transitions and other parts of the model and uses this information to assign higher priorities to the test cases. EFSM models consist of two types of dependences: control and data dependences. The results show that model dependence-based test prioritization (considering only two types of dependences) gives improvement in the effectiveness of test prioritization. The corresponding system for each model was implemented in C language. In another work, Korel et al. [43] compared the effectiveness of different prioritization heuristics. The results show that model based prioritization along with heuristic 5 gave the best performance. Heuristic 5 states that each modified transition should have the same opportunity of getting executed by the test cases. Korel et al. [44] proposed another approach of prioritization using the heuristics discussed in [43]. In this new approach, they considered the changes made in the source code and identified the elements of the model that are related to these changes to prioritize the test cases. In our approach, we have considered the object-oriented programs in Java. The program is represented by our proposed intermediate graph. The graph is constructed by considering many more dependences that exist among the program parts in addition to control and data dependences, giving a clear visualization of the dependences. Then, we identify the effect of modifications and represent the affected program parts in another graph. Our representation is more adaptable to the frequent changes of the software and our approach relies on the execution of these affected program parts. Thus, our prioritization approach is based on both the coverage of the affected program parts and the fault exposing potential of the test cases.

Jeffrey and Gupta [45] proposed a prioritization approach using the relevant slices. They also aimed for early detection of faults during regression testing process. This approach considers the execution of the modified statements for prioritizing the test cases. The assumption is that if any modification results in some faulty output for a test case, then it must affect some computation in the relevant slice of that test case. Therefore, the test case having higher number of statements is given higher priority assuming that they have a better potential to expose the faults. However, intuitively, not all statements depending upon some modification will have the same level of fault-proneness. It may so happen that a test case executing less number of statements will detect more faults than another test case that executed more number of statements. The level of fault-proneness of the statements executed by the test cases affects the fault exposing potential of that test case. Therefore, in our approach, we computed the coupling values of the affected program parts to identify the probable fault-proneness of these programs parts. Our approach assigns a higher priority to that test case which executes maximum number of high fault-prone statements. Further, unlike our hierarchical decomposition slicing approach, relevant slicing depends upon the execution trace of the test cases and is proposed to work on C programs. Even though execution trace based slicing would result in slices of smaller sizes, the computational overhead is very high. The efficiency of our slicing approach is shown in Table 2. We have also shown the time requirement of our prioritization approach in Table 5.

The performance goal of the prioritization approach proposed by Kayes [11] is based on how quickly the dependences among the faults are identified in the regression testing process. An early detection of the fault dependences would enable faster debugging of the faults. The paper assumes that the knowledge of the fault presence is extracted from the previous executions of the test cases. A fault dependence graph is constructed using this information. However, one major limitation of this approach is that regression testing aims at discovering new faults introduced by the changes made to the software. But the prioritization approach proposed in this paper only enhances the chances of finding the faults which have already been revealed and are present in the fault dependence graph. New faults if any cannot be discovered. Further, this approach does not take into account the fault-proneness of the statements. However, our approach relies on the dependence of the affected program parts represented as affected slice graph (ASG), so that error propagation because of the change is better visualized and analyzed. We compute the fault-proneness of the statements by computing their coupling values as coupling measures are proven to be good indicator of fault-proneness. Thus, our approach has a higher probability of exposing new faults, if any, in the software.

Mei et al. [15] proposed a static prioritization technique to prioritize the JUnit test cases. This prioritization technique is independent of the coverage information of the test cases. It works on the analysis of the static call graphs of JUnit test cases and the program under test to estimate the ability of each test case to achieve code coverage. The test cases are scheduled based on these estimates. The experiments are carried out on 19 versions of four Java programs of considerable size considering their method and class level JUnit test cases. The heuristic to prioritize the test cases in this approach is to cover system components (in terms of total components covered or components newly covered). The coverage of the system components acts as a proxy for evaluating a test cases true potential of exposing faults. If any two test cases carry the same heuristic value then the approach randomly decides on the test case to be given higher priority. Though this is a scalable approach as it works at coarse granularity level and incurs less computational cost, it suffers from many limitations. The prioritization techniques that work at a finer granularity level give better performances (in terms of fault exposing potential) as compared to the techniques that work at coarse granularity level [42]. This approach ignores the faults caused by many object-oriented features such as inheritance, polymorphism, and dynamic binding and focuses only on the static call relationships of the methods in the form of a call graph. Static call relationships are more to procedure-oriented programs. Interaction and communication between methods in the form of message passing is highly important in object-oriented programs. A single method is invoked by different objects and the behavior of the method also differs accordingly. Any prioritization technique is efficient if it is based on the characteristics of the program to be tested. Therefore, considering the object-oriented features is essential. Java supports encapsulation and provides four access levels (private, public, protected, and default) to access the data members and member methods. Any misinterpretation of these access levels forms a rich source of faults. Java supports a feature named "super" to have access to the base class constructor from the derived class constructor. This additional dependence between constructors of the derived class and the bases class needs attention of the testers. Method overriding allows a method in the derived class to have the same function signature as the method in its parent. If invocation to such methods is not resolved correctly, then it can cause some serious faults. Another powerful feature and a potential source of fault is variable hiding. It allows declaration of a variable with the same name and type in the derived class as it is in the base class and allows both variables to reside in the derived class. Problem arises when an incorrect variable is accessed. Inheritance is a powerful feature but sometimes unintentional misuse of this feature can result in serious faults. Polymorphism in Java exists for both attributes and methods and both use dynamic binding. An object of its class type can access an attribute or method of its subclass type. The subclass object can also access the same attributes and methods. These attributes and methods behave differently depending upon the kind of object that is referring it. Such polymorphic dependences if not resolved can cause faults. Interested readers are requested to refer to [46–49] for more number of faults introduced by the misuse of the object-oriented features. Therefore, any prioritization technique with a performance goal of revealing more faults must consider the object-oriented features as they can induce many kinds of faults in the system. Our approach considers all the object-oriented features in the form of intermediate graph. This intermediate graph is constructed by identifying the dependences that can exist among various program parts and are given in [32]. Our approach works at a finer granularity level and therefore may not be as scalable as [15] but has better fault exposing potential.

Fang et al. [14] have proposed similarity based prioritization technique. The authors have taken five Java programs from Software artifacts Infrastructure Repository (SIR) [37] to validate their approach. The prioritization process is based on the ordered sequence of the program entities. They propose two algorithms farthest-first ordered sequence (FOS) and greed-aided clustering ordered sequence (GOS). The FOS approach first selects the test case having largest statement coverage. The next test case that is selected is the one that is farthest in distance from the already selected test case. It computes two types of distances: a pairwise distance between the test cases and distance between a candidate test case and the already selected ones. GOS approach consists of clusters of test cases in which initially each cluster consists of only one test case. Then the clusters are merged depending upon the minimum distance between any two clusters. This process of merging the clusters is repeated until the size of the cluster set is less than some given $n$. Then, the algorithm iteratively chooses one test case from each cluster and adds to the prioritized test suite until all the clusters are empty. The experimental results in this study show that statement coverage is most efficient and preferred for prioritization. When the size of the test suite is large, then additional measures are taken to reduce the cost of prioritization. This approach gives equal importance to all the test cases assuming that all the test cases have equal potential of exposing the faults. Intuitively, a test case executing less number of statements can expose more faults provided that the covered statements have high proneness to faults. It also does not consider the object-oriented features and the faults generated by these features. Unlike Fang et al. [14], we consider the fault inducing capability of the object-oriented features based on which we detect the affected program parts. We propose to prioritize a set of *change-based selected* test cases that are relevant to validate the change under regression testing. We compute the fault-proneness of the affected statements and then prioritize the test cases based on the coverage of these high fault-prone statements (represented as nodes in our proposed graph).

Lou et al. [16] proposed a mutation-based prioritization technique. In this approach, they compared the two versions of the same software to find the modification. Then, they generate the mutants only for the modified code. They selected only those test cases of the original version that worked on the new version of the software for prioritization. The test case that killed more mutants was given higher priority. The authors used a mutation generation tool, named *Javalanche*. Unlike our approach, Lou et al. [16] do not take into consideration the object-oriented features and the faults likely to occur because of these features. It is also silent on the type of mutation operators (faults) considered for their experimentation. Like Lou et al. [16], we generate mutants only for the sliced program (representing the affected program parts). However, we used MuClipse (an eclipse version of MuJava) to generate the mutation faults. We use coupling measure of the affected program parts as a surrogate to imply fault-proneness. Our hypothesis assumes that the test cases that execute the nodes with high coupling value have a higher chance of detecting faults early during regression testing. We used mutation analysis only to validate our hypothesis.

The detail survey conducted on available coverage based prioritization techniques [11, 14–16, 41–45] reveals that these techniques have not considered the object-oriented features. The presence of many faults arising due to different object-oriented features is inherent to object-oriented programs and hence must be considered. Therefore, we find that the approaches contributed to by Panigrahi and Mall [17, 18] relate closely to our approach for an experimental comparison. Panigrahi and Mall proposed a version specific prioritization technique [17] to prioritize the test cases of object-oriented programs. Their technique prioritizes the selected regression test cases. The test cases are prioritized based on the coverage of affected nodes of an intermediate graph model of the program under consideration. The affected nodes are determined due to the dependences arising on account of the object relations in addition to the data and control dependences. The effectiveness of their approach is shown in form of improved APFD measure achieved for the test cases. In another work, Panigrahi and Mall [18] have improved their earlier work [17] by achieving a better APFD value. In this technique, the affected nodes are initially assigned a weight of 1. The weight is decreased by 0.5, whenever that node is covered by previous execution of the test cases. In both approaches [17, 18], they have assumed that all the test cases have equal cost, and all faults have the same severity. The assumption is also that all the affected nodes have a uniform distribution of faults. As a result, a test case executing more number of affected nodes will detect more faults and, therefore, has a higher priority. The average percentage of affected nodes covered by this approach is shown in Figure 7. Unlike the approach in [18] that is based on node coverage only, our proposed approach is based on the fact that some nodes are more fault-prone than other nodes. We used an intermediate graph that represents only those nodes that are affected by the modification made to the program to compute the fault-proneness of the nodes. The coupling factor of each node in the ASG is computed to predict its level of fault-proneness. The test cases are then prioritized based on the fault-prone nodes that they execute. Unlike [18], a test case executing more number of fault-prone nodes has a higher computed weight and gets a higher priority in our approach.

*5.1. Threats to Validity.* It is obvious for any new proposed work to be associated with some threat to its validity, and it is likely for this work as well. Our approach is capable of measuring the coupling value of a class in the presence of many object-oriented features such as inheritance, interfaces, polymorphism, and templates. The coupling of classes in a *subclass-superclass* relationship can have a different impact on software maintainability and fault-proneness compared to the coupling of classes that are not in such relationship. Therefore, it is essential to make a distinction between coupling within an inheritance hierarchy and coupling across inheritance hierarchies. Similarly, whether the presence of *Java interfaces* (that usually do not contain actual implementations) contributes to the coupling measurement or not is a matter of study that is not included in this paper. The impact of *inclusion/exclusion* of any of the object-oriented features on the coupling measurement has not been empirically investigated in this paper. We believe that a detailed empirical research on such relationships and their impact on the proposed coupling analysis is essential and is left for future study. Another threat to the validity of this work is that the fault prediction can be improved when both coupling and cohesion metrics are considered together [20], but this approach focuses only on coupling measure. Slicing techniques based on intermediate graphs are always limited by the scalability issues of the graph for larger program. This approach is tested to work well with programs having nearly 1 Lakh line of code. For larger programs it may raise some memory issues. However, it will work fine for bigger programs if the graph is restricted to method level analysis only. The limited size and complexity of the experimental programs are considered a threat to the validity of this approach. Our approach considers only the primary mutants. It does not consider the secondary mutants which are also important. Our approach of mutation analysis may be extended to handle secondary mutants. The use of mutation analysis for the fault manipulation of these programs may not represent the actual fault occurrence in the complex industrial programs and hence is considered a threat to this approach. Though the proposed prioritization approach is efficient in detecting the faults, it may not be so in terms of time requirement. However, the time requirement is within acceptable limit if applied to the test cases selected for regression testing, and the coverage information is available. An empirical study of the impact of prioritization time on the choice of selection of the prioritization techniques would be interesting and may be carried out in future.

## 6. Conclusion and Future Work

In this paper, we proposed a coupling metric based technique to improve the effectiveness of test case prioritization in regression testing. Analysis is done to show that prioritized test cases are more effective in exposing the faults early in the regression test cycle. We performed hierarchical decomposition slicing on the intermediate graph of the input program. The affected component coupling (ACC) value of each node of the ASG is calculated as a measure to predict its fault-proneness. In this technique, weight is assigned to each node of ASG based on its ACC value. The weight of a test case in a given test suite is then calculated by adding the weights of all the nodes covered by it. The test cases are prioritized based on their coverage of fault-prone affected nodes. Thus, the test case with a higher weight is given higher priority in the test suite. The results show that our FPANC approach achieves approximately 8% of increase in the APFD metric value over ANC approach. In the future, we aim to prioritize the test cases for more complex object-oriented (OO) programs such as concurrent and distributed OO programs. We would also like to incorporate different other coupling measures and metrics to predict the fault-proneness of modules and prioritize the test cases based on their coverage weights. We as well aim to compute the cohesion values of the program elements and use them along with their coupling values for a better fault prediction analysis and prioritization.

## Competing Interests

The authors declare that they have no competing interests.